\documentclass[a4paper,11pt]{article}
\usepackage{amssymb}
\usepackage{amsmath}
\usepackage{epsfig}
\usepackage{subfigure}
\usepackage{graphics}
\usepackage{tikz}
\usepackage{latexsym}
\usepackage{rotating}
\usepackage{titlesec}
\usetikzlibrary{matrix}

\title{$(2+1)$-dimensional AKNS($-N$) Systems:\, $ N=3,4$}

\author{Metin G\"{u}rses \thanks{gurses@fen.bilkent.edu.tr}\\
{\small Department of Mathematics, Faculty of Science}\\
{\small Bilkent University, 06800 Ankara - Turkey}\\
Asl{\i} Pekcan \thanks{Email:aslipekcan@hacettepe.edu.tr} \\
{\small Department of Mathematics, Faculty of Science} \\
{\small Hacettepe University, 06800 Ankara - Turkey}
}

\date{\nonumber}
\begin{document}
\maketitle
\date{\nonumber}
\begin{abstract}
In this work we continue to study negative AKNS($N$) that is AKNS($-N$) system for $N=3,4$. We obtain all possible local and nonlocal reductions of these equations.  We construct the Hirota bilinear forms of these equations and find one-soliton solutions. From the reduction formulas we obtain also one-soliton solutions of all reduced equations.
\end{abstract}

\section{Introduction}

The AKNS hierarchy \cite{AKNS} is

\begin{equation}\label{generalsys}
cu_{t_N}=\mathcal{R}^{N} u_x \,\, \mathrm{where} \,\,  u= \left( \begin{array}{c}
p  \\
q
 \end{array} \right) \,\, \mathrm{i.e.} \,\,  \left( \begin{array}{c}
p_{t_N}  \\
q_{t_N}

 \end{array} \right)= \mathcal{R}^{N} \left( \begin{array}{c}
p_x  \\
q_x
 \end{array} \right),
\end{equation}
\noindent for $N=0,1,2,\cdots$, where $\mathcal{R}$ is the recursion operator,
\begin{equation}\label{recursion}
 \mathcal{R}=\left( \begin{array}{cc}
-pD^{-1}q+\frac{1}{2}D & -pD^{-1}p  \\
qD^{-1}q & qD^{-1}p-\frac{1}{2}D
 \end{array} \right)
\end{equation}
\noindent and $c$ is an arbitrary constant. Here $D$ is the total $x$-derivative and $D^{-1}=\int^x$ is the standard anti-derivative.
In the last decade there have been considerable amount of works on nonlocal reductions of this system. Nonlinear Schr\"{o}dinger (NLS) system ($N=1$) and its reductions \cite{gerd}-\cite{origin}, multidimensional versions of NLS and their reductions \cite{fok}-\cite{ZL}, modified KdV (mKdV) system ($N=2$) and its reductions \cite{AbMu3}, \cite{chen}, \cite{GurPek3}-\cite{origin}, \cite{GurPek2}-\cite{Pek} have attracted many researchers to obtain interesting soliton solutions by using inverse scattering method, Darboux transformations, and Hirota bilinear method. There are also some interests on higher members of the system (\ref{generalsys}) for $N\ge 3$ \cite{fourthNLS1}-\cite{sixthNLS3}. These methods, in particular the Hirota method, become more difficult or not available for systems $N \ge 3$. Similar difficulties arise also for negative AKNS hierarchy in $(2+1)$-dimensions \cite{GurPek4}.

Letting $p=\frac{g}{f}$, $q=\frac{h}{f}$, and $t_N\rightarrow -\frac{t_N}{2^{N}}$ then the coupled partial differential equations for $N=0,1,2$ have similar Hirota bilinear forms
\begin{align}
&(cD_{t_N}+D_x^{N+1})\{g\cdot f\}=0, \label{aknshirota1}\\
&(cD_{t_N}+(-1)^{N}\,D_x^{N+1})\{h\cdot f\}=0, \label{aknshirota2}\\
&D_x^2\{f\cdot f\}=-2gh. \label{aknshirota3}
\end{align}

In our previous work \cite{GurPek4} (see also \cite{gurpek}) we started to investigate $(2+1)$-dimensional negative AKNS system, namely AKNS($-N$) system defined as
\begin{equation}\label{AKNSextension}
\mathcal{R}(u_{t_N})-a\mathcal{R}^{N} (u_{x})=b\,u_y \ \mathrm{for} \ N=0,1,2,\cdots,
\end{equation}
 Here $a$ and $b$ are any constants. In \cite{GurPek4} we have considered only the cases for $N\leq 2$, and constructed the Hirota bilinear forms only for these negative AKNS systems. We obtained soliton solutions, and found all local and
 nonlocal reductions of these systems. Similar to AKNS($N$) systems by letting
\begin{equation}\label{hir2}
 p=\frac{g}{f}, \quad q=\frac{h}{f}, \quad \frac{gh}{f^2}=- \left(\frac{f_{x}}{f} \right)_{x},
 \end{equation}
 we obtain the Hirota bilinear form of $(2+1)$-dimensional AKNS($-N$) system as \cite{GurPek4}
  \begin{eqnarray}
&& (b D_{y}-\frac{1}{2}\, D_{x} D_{t_N}+\frac{a}{2^N}\,D_{x}^{N+1} )\{ g \cdot f\}=0, \label{negn=0hirota1}\\
&&  (b D_{y}+\frac{1}{2}\, D_{x} D_{t_N}+(-1)^N\,\frac{a}{2^N}\, D_{x}^{N+1}) \{ h \cdot f\}=0,\label{negn=0hirota2}\\
&&  D_{x}^2 \{f \cdot f\}=-2 g h. \label{negn=0hirota3}
\end{eqnarray}
for $N=0,1,2$.

The Hirota bilinear forms (\ref{aknshirota1})-(\ref{aknshirota3}) of the AKNS($N$) system are  simpler and can be written in a nice form  for $N=0,1,2$. Due to this property of the Hirota bilinear forms for AKNS($N$) system for $N=0,1,2$
the Hirota bilinear forms (\ref{negn=0hirota1})-(\ref{negn=0hirota3}) of the negative AKNS system has also similar nice structure corresponding to $N=0, 1, 2$. This simplicity is valid only for $N=0,1,2$. For $N \ge 3$ the Hirota bilinear forms of the system were unknown and due to this fact we could not study the Hirota bilinear forms and soliton solutions of negative AKNS systems in \cite{GurPek4}.

Recently \cite{gurpek1} we have succeeded to construct the Hirota bilinear forms of AKNS($N$) system for $N=3,4,5,6$ and discussed several problems arising in obtaining Hirota bilinear forms of higher members  of the system. The Hirota bilinear forms of AKNS($N$) system  for $N \ge 3$ start to  contain inhomogeneous terms. These terms are relatively simpler for $N=3$ and $N=4$. Hence to construct the Hirota bilinear forms of the systems AKNS(-3) and AKNS(-4) are now practically possible.

In this work we construct first the Hirota bilinear  forms of AKNS(-3) and AKNS(-4) systems. In Section 3 we obtain one-soliton solutions of these two systems. We present local and nonlocal reductions of these negative AKNS systems in Sections 4 and 5 respectively. In Section 6 we give one-soliton solutions of these reduced nonlocal equations.

\section{Hirota bilinear forms of $(2+1)$-dimensional AKNS(-N) systems for $N=3, 4$}

\noindent In this  section we study the $(2+1)$-dimensional AKNS(-3) and AKNS(-4) systems and obtain the Hirota bilinear forms of them.
For $N=3$ the system (\ref{AKNSextension}) gives the $(2+1)$-dimensional AKNS(-3) system as
\begin{align}
&bp_y=\frac{1}{2}p_{xt_3}-\frac{a}{8}p_{xxxx}+\frac{3a}{4}qp_x^2+\frac{a}{2}pp_xq_x+apqp_{xx}+\frac{a}{4}p^2q_{xx}-\frac{3a}{4}p^3q^2-pD^{-1}(pq)_{t_3},\label{negn=3p_y}\\
&bq_y=-\frac{1}{2}q_{xt_3}+\frac{a}{8}q_{xxxx}-\frac{3a}{4}pq_x^2-\frac{a}{2}qq_xp_x-aqpq_{xx}-\frac{a}{4}q^2p_{xx}+\frac{3a}{4}p^2q^3+qD^{-1}(pq)_{t_3}.\label{negn=3q_y}
\end{align}
If we use (\ref{hir2}) in the above system, we obtain the Hirota bilinear form of the
$(2+1)$-dimensional AKNS(-3) system as
\begin{align}
&(bD_y-\frac{1}{2}D_xD_{t_3}+\frac{a}{8}D_x^4)\{g\cdot f\}=-\frac{3}{8}ahs,\label{negn=3-1}\\
&(bD_y+\frac{1}{2}D_xD_{t_3}-\frac{a}{8}D_x^4)\{h\cdot f\}=\frac{3}{8}ag\tau,\label{negn=3-2}\\
& D_x^2\{f\cdot f \}=-2gh ,\label{negn=3-3}\\
& D_x^2\{g\cdot g\}=fs,\label{negn=3-4}\\
&D_x^2\{h\cdot h\}=f\tau,\label{negn=3-5}
\end{align}
where $s$ and $\tau$ are auxiliary functions.\\

\noindent For $N=4$ the system (\ref{AKNSextension}) gives the $(2+1)$-dimensional AKNS(-4) system as
\begin{align}
bp_y=&\frac{1}{2}p_{xt_4}-\frac{a}{16}p_{5x}+\frac{5a}{8}pqp_{xxx}+\frac{5a}{8}pp_xq_{xx}-\frac{15a}{8}p^2q^2p_x+\frac{5a}{8}pp_{xx}q_x+\frac{5a}{4}qp_xp_{xx}\nonumber\\
 &+\frac{5a}{8}p_x^2q_x-pD^{-1}(pq)_t\label{negn=4p_y}\\
bq_y=&-\frac{1}{2}q_{xt_4}-\frac{a}{16}q_{5x}+\frac{5a}{8}qpq_{xxx}+\frac{5a}{8}qq_xp_{xx}-\frac{15a}{8}q^2p^2q_x+\frac{5a}{8}qq_{xx}p_x+\frac{5a}{4}pq_xq_{xx}\nonumber\\
 &+\frac{5a}{8}q_x^2p_x+qD^{-1}(pq)_t.\label{negn=4q_y}
\end{align}
By using (\ref{hir2}) in the system (\ref{negn=4p_y}) and (\ref{negn=4q_y}) we get the Hirota bilinear form of the
$(2+1)$-dimensional AKNS(-4) system as
\begin{align}
&(bD_y-\frac{1}{2}D_xD_{t_4}+\frac{a}{16}D_x^5)\{g\cdot f\}=\frac{5}{16}aD_x\{h\cdot s\},\label{negn=4-1}\\
&(bD_y+\frac{1}{2}D_xD_{t_4}+\frac{a}{16}D_x^5)\{h\cdot f\}=\frac{5}{16}aD_x\{g\cdot \tau\},\label{negn=4-2}\\
& D_x^2\{f\cdot f \}=-2gh ,\label{negn=4-3}\\
& D_x^2\{g\cdot g\}=fs,\label{negn=4-4}\\
&D_x^2\{h\cdot h\}=f\tau,\label{negn=4-5}
\end{align}
where $s$ and $\tau$ are auxiliary functions.

\section{One-soliton solutions $(2+1)$-dimensional AKNS(-N) systems for $N=3, 4$}

To obtain one-soliton solutions of AKNS(-3) and AKNS(-4) we take $g=\varepsilon g_1$, $h=\varepsilon h_1$, $f=1+\varepsilon^2 f_2$, $s=\varepsilon^2 s_2$, and $\tau=\varepsilon^2 \tau_2$, where
\begin{equation}
g_1=e^{\theta_1}, \quad h_1=e^{\theta_2}
\end{equation}
for $\theta_i=k_ix+\rho_iy+\omega_it+\delta_i$, $i=1, 2$. For simplicity, we use $t$ for both $t_3$ and $t_4$. We insert these expansions in the
Hirota bilinear forms of AKNS(-3) and AKNS(-4) systems and consider the coefficients of $\varepsilon^n$, $n=1,2,3,4$. The coefficients of $\varepsilon$ give the dispersion
relations as
\begin{equation}\label{dispersionAKNS(-3)}
\rho_1=\frac{1}{b}(\frac{1}{2}k_1\omega_1-\frac{a}{8}k_1^4),\quad \rho_2=\frac{1}{b}(-\frac{1}{2}k_2\omega_2+\frac{a}{8}k_2^4)
\end{equation}
for AKNS(-3) system, and
\begin{equation}\label{dispersionAKNS(-4)}
 \rho_1=\frac{1}{b}(\frac{1}{2}k_1\omega_1-\frac{a}{16}k_1^5), \quad \rho_2=\frac{1}{b}(-\frac{1}{2}k_2\omega_2-\frac{a}{16}k_2^5)
\end{equation}
for AKNS(-4) system. From the coefficients of $\varepsilon^2$ we have
\begin{equation}\displaystyle
f_2=-\frac{e^{\theta_1+\theta_2}}{(k_1+k_2)^2},
\end{equation}
and $s_2=\tau_2=0$. The coefficients of $\varepsilon^3$ and $\varepsilon^4$ vanish directly.

\noindent Let us take $\varepsilon=1$. Hence one-soliton solutions of the AKNS(-3) system given by (\ref{negn=3p_y}) and (\ref{negn=3q_y}) and the AKNS(-4) system given by (\ref{negn=4p_y}) and (\ref{negn=4q_y}) are $(p(x,y,t), q(x,y,t))$ where
\begin{equation}\label{one-sol}
p(x,y,t)=\frac{e^{\theta_1}}{1-\frac{e^{\theta_1+\theta_2}}{(k_1+k_2)^2}},\quad
q(x,y,t)=\frac{e^{\theta_2}}{1-\frac{e^{\theta_1+\theta_2}}{(k_1+k_2)^2}}
\end{equation}
for $\theta_i=k_ix+\rho_iy+\omega_it+\delta_i$, $i=1, 2$, where the dispersion relations are given in (\ref{dispersionAKNS(-3)}) and
(\ref{dispersionAKNS(-4)}), respectively. Here $k_i, \delta_i$, $i=1, 2$ are arbitrary complex numbers.\\

\noindent By using the one-soliton solutions of the $(2+1)$-dimensional AKNS($-N$) systems for $N=3, 4$  we can also obtain soliton
solutions of the local and nonlocal reductions of these systems. Note that here in our solutions we will focus on the domain $t\geq 0$, $x\in \mathbb{R}$.
Before presenting the soliton solutions of the nonlocal (and local) reduced $(2+1)$-dimensional AKNS($-N$) equations for $N=3, 4$, we will first give local and nonlocal consistent reductions.

\section{Local reductions of $(2+1)$-dimensional AKNS($-N$) systems for $N=3, 4$}

In this section we present local reductions of $(2+1)$-dimensional AKNS(-3) and AKNS(-4) systems. We first obtain the constraints to have these local reductions consistently. Then we give reduced local equations under these constraints. Note that both AKNS(-3) and AKNS(-4) systems do not possess the local reduction $q(x,y,t)=k p(x,y,t)$, $k$ is a real constant.

\subsection{$q(x,y,t)=k\bar{p}(x,y,t)$, $k$ is a real constant}

\noindent Under this reduction the AKNS(-3) system (\ref{negn=3p_y}) and (\ref{negn=3q_y})
reduces to the local complex AKNS(-3) equation
\begin{equation}\label{localcomplexAKNS(-3)}
bp_y=\frac{1}{2}p_{xt}-\frac{a}{8}p_{xxxx}+\frac{3a}{4}k\bar{p}p_x^2+\frac{a}{2}kp|p_x|^2+ak|p|^2p_{xx}+\frac{a}{4}kp^2\bar{p}_{xx}-\frac{3a}{4}k^2p|p|^4-kpD^{-1}(|p|^2)_t,
\end{equation}
where $\bar{b}=-b$, $\bar{a}=a$, consistently. \\

\noindent When we apply this reduction to the AKNS(-4) system (\ref{negn=4p_y}) and (\ref{negn=4q_y}) the system
consistently reduces to the local complex AKNS(-4) equation
\begin{align}\label{localcomplexAKNS(-4)}
bp_y=&\frac{1}{2}p_{xt}-\frac{a}{16}p_{5x}+\frac{5a}{8}kpp_x\bar{p}_{xx}-\frac{15a}{8}k^2|p|^4p_x+\frac{5a}{8}kp\bar{p}_xp_{xx}+\frac{5a}{4}k\bar{p}p_xp_{xx}
+\frac{5a}{8}k|p_x|^2p_x^2\nonumber\\
&-kpD^{-1}(|p|^2)_t,
\end{align}
where $\bar{b}=-b$, $\bar{a}=-a$.

\section{Nonlocal reductions of $(2+1)$-dimensional AKNS($-N$) systems for $N=3, 4$}

In order to have consistent nonlocal reductions we use the following representation for $D^{-1}$ \cite{GurPek4}
\begin{equation}
D^{-1}\,F=\frac{1}{2}\, \left(\int_{-\infty}^{x}-\int_{x}^{\infty}\,\right) F(x^{\prime},y,t) dx^{\prime}.
\end{equation}
We define $\xi(x,y,t)$ which is invariant under the discrete transformations $x \to \varepsilon_{1} x$, $y \to \varepsilon_{2} y,$ and $t \to \varepsilon_{3} t$ as
\begin{equation}
\xi (x,y,t)=D^{-1}\,p\, p^{\varepsilon} \equiv \left(\int_{-\infty}^{x}-\int_{x}^{\infty}\,\right) p(x^{\prime},y,t)\, p(\varepsilon_{1}\, x^{\prime}, \varepsilon_{2}\, y, \varepsilon_{3}\, t)\, dx^{\prime},
\end{equation}
where $\varepsilon_{j}^2=1, j=1,2,3$. It is clear that
\begin{equation}
\xi(\varepsilon_{1}\, x, \varepsilon_{2}\, y, \varepsilon_{3}\, t)=\varepsilon_{1}\, \xi(x,y,t).
\end{equation}

In this following part we present nonlocal reductions of $(2+1)$-dimensional AKNS(-3) and AKNS(-4) systems. We obtain the conditions on the parameters to have these nonlocal reductions consistently. Then we give reduced nonlocal equations under these conditions.

\subsection{$q(x,y,t)=kp(\varepsilon_1x,\varepsilon_2y, \varepsilon_3t)$, $\varepsilon_1^2=\varepsilon_2^2=\varepsilon_3^2=1$, $k$ is a real constant}

\noindent Under this reduction the AKNS(-3) system given by (\ref{negn=3p_y}) and (\ref{negn=3q_y}) reduces consistently to the following nonlocal AKNS(-3) equation:
\begin{equation}
b p_{y}=\frac{1}{2}p_{tx}-\frac{a}{8} p_{xxxx}+\frac{3a}{4}kp^{\varepsilon}p_x^2+\frac{a}{2}kpp_xp_x^{\varepsilon}+akpp^{\varepsilon}p_{xx}
   +\frac{a}{4}kp^2p_{xx}^{\varepsilon}-\frac{3a}{4}k^2p^3(p^{\varepsilon})^2-k p D^{-1} (p p^{\epsilon})_{t},\label{realnonlocalredn=3}
\end{equation}
where $k$ is any real constant. Here $p^{\epsilon}= p(\epsilon_{1} x,\epsilon_{2} y, \epsilon_{3} t)$. The above equation is valid only when $\epsilon_{2}=-1$ and $\epsilon_{1} \epsilon_{3}=1$. Therefore we have only $p^{\epsilon}=p(x,-y,t)$ and $p^{\epsilon}=p(-x,-y,-t)$ cases as for the AKNS(-1) system \cite{GurPek4}.\\

\noindent When we apply this reduction to the AKNS(-4) system (\ref{negn=4p_y}) and (\ref{negn=4q_y}), it reduces consistently to the following nonlocal AKNS(-4) equation:
\begin{align}
bp_{y}=&\frac{1}{2}p_{tx}-\frac{a}{16}p_{5x}+\frac{5a}{8}kpp^{\varepsilon}p_{xxx}+\frac{5a}{8}kpp_xp_{xx}^{\varepsilon}-\frac{15a}{8}k^2p^2(p^{\varepsilon})^2p_x
+\frac{5a}{8}kpp_{xx}p_x^{\varepsilon}+\frac{5a}{4}kp^{\varepsilon}p_xp_{xx}\nonumber\\
&+\frac{5a}{8}kp_x^2p_x^{\varepsilon}-kpD^{-1}(pp^{\epsilon})_{t},\label{realnonlocalredn=4}
\end{align}
where $k$ is any real constant and $p^{\epsilon}= p(\epsilon_{1} x,\epsilon_{2} y, \epsilon_{3} t)$. This equation is valid only when $\epsilon_{3}=-1$ and $\epsilon_{1}\, \epsilon_{2}=1$. Here we have also two possible cases; $p^{\epsilon}=p(x,y,-t)$ and $p^{\epsilon}=p(-x,-y,-t)$ as for the AKNS(0) and AKNS(-2) systems \cite{GurPek4}.

\subsection{$q(x,y,t)=k\bar{p}(\varepsilon_1x,\varepsilon_2y, \varepsilon_3t)$, $\varepsilon_1^2=\varepsilon_2^2=\varepsilon_3^2=1$, $k$ is a real constant}

\noindent  When we apply this reduction to the AKNS(-3) system (\ref{negn=3p_y}) and (\ref{negn=3q_y}), it reduces consistently to the nonlocal complex AKNS(-3) equation,
\begin{equation}
b p_{y}=\frac{1}{2}p_{tx}-\frac{a}{8} p_{xxxx}+\frac{3a}{4}k\bar{p}^{\varepsilon}p_x^2+\frac{a}{2}kpp_x\bar{p}_x^{\varepsilon}+akp\bar{p}^{\varepsilon}p_{xx}
+\frac{a}{4}kp^2\bar{p}_{xx}^{\varepsilon}-\frac{3a}{4}k^2p^3(\bar{p}^{\varepsilon})^2-k p D^{-1} (p \bar{p}^{\epsilon})_{t}, \label{complexnonlocalredn=3}
\end{equation}
where $k$ is any real constant, if the following constraints hold:
\begin{equation}\label{complexconstn=3}
\epsilon_{1}\, \epsilon_{2}\, \epsilon_{3}\, \bar{b}=-b, \quad \epsilon_{1}\, \epsilon_{3}\, \bar{a}=a.
\end{equation}
Therefore we have seven different time and space reversals as in the AKNS(-1) case \cite{GurPek4}:

\vspace{0.2cm}
\noindent
(i) $p^{\epsilon}(x,y,t)=p(-x,y,t)$, where $a$ is pure imaginary and $b$ is real.\\
(ii) $p^{\epsilon}(x,y,t)=p(x,-y,t)$, where $a$ and $b$ are real.\\
(iii) $p^{\epsilon}(x,y,t)=p(x,y,-t)$, where $a$ is pure imaginary and $b$ are real.\\
(iv) $p^{\epsilon}(x,y,t)=p(-x,-y,t)$, where $a$ and $b$ are pure imaginary. \\
(v) $p^{\epsilon}(x,y,t)=p(-x,y,-t)$, where $a$ is real and $b$ is pure imaginary.\\
(vi) $p^{\epsilon}(x,y,t)=p(x,-y,-t)$, where $a$ and  $b$ are pure imaginary. \\
(vii) $p^{\epsilon}(x,y,t)=p(-x,-y,-t)$, where $a$ and $b$ are real.\\

\noindent Each case (i)-(vii) gives a $(2+1)$-dimensional nonlocal equation in the form of (\ref{complexnonlocalredn=3}).

\noindent Under this reduction the AKNS(-4) system (\ref{negn=3p_y}) and (\ref{negn=3q_y}) reduces consistently to the nonlocal complex AKNS(-4) equation,
\begin{align}
b p_{y}=&\frac{1}{2}p_{tx}-\frac{a}{16} p_{5x}+\frac{5a}{8}kp\bar{p}^{\varepsilon}p_{xxx}+\frac{5a}{8}kpp_x\bar{p}_{xx}^{\varepsilon}-\frac{15a}{8}k^2p^2(\bar{p}^{\varepsilon})^2p_x
+\frac{5a}{8}kpp_{xx}\bar{p}_x^{\varepsilon}+\frac{5a}{4}k\bar{p}^{\varepsilon}p_xp_{xx}\nonumber\\
&+\frac{5a}{8}kp_x^2\bar{p}_x^{\varepsilon}-k p D^{-1} (p \bar{p}^{\epsilon})_{t},\label{complexnonlocalredn=4}
\end{align}
where $k$ is any real constant, if the following constraints hold:
\begin{equation}\label{complexconstn=4}
\epsilon_{1} \epsilon_{2} \epsilon_{3} \bar{b}=-b, \quad \epsilon_{3} \bar{a}=-a.
\end{equation}
Therefore we have seven different time and space reversals as in the AKNS(0) and AKNS(-2) cases \cite{GurPek4}:

\vspace{0.2cm}
\noindent
(i) $p^{\epsilon}(x,y,t)=p(-x,y,t)$, where $a$ is pure imaginary and $b$ is real.\\
(ii) $p^{\epsilon}(x,y,t)=p(x,-y,t)$, where $a$ is pure imaginary and $b$ is real.\\
(iii) $p^{\epsilon}(x,y,t)=p(x,y,-t)$, where $a$ and $b$ are real.\\
(iv) $p^{\epsilon}(x,y,t)=p(-x,-y,t)$, where $a$ and $b$ are pure imaginary. \\
(v) $p^{\epsilon}(x,y,t)=p(-x,y,-t)$, where $a$ is real and $b$ is pure imaginary.\\
(vi) $p^{\epsilon}(x,y,t)=p(x,-y,-t)$, where $a$ is real and  $b$ is pure imaginary. \\
(vii) $p^{\epsilon}(x,y,t)=p(-x,-y,-t)$, where $a$ and $b$ are real.\\

\noindent
Each case (i)-(vii) gives a $(2+1)$-dimensional nonlocal equation in the form of (\ref{complexnonlocalredn=4}).

\section{One-soliton solutions of local and nonlocal reduced complex AKNS(-3) and AKNS(-4) equations}
In this part we first give the constraints on the parameters of one-soliton solutions of the reduced local and nonlocal equations.
Then we present one-soliton solutions obeying these constraints.

\subsection{One-soliton solutions of local reduced equations}

Let us use the local reduction formula $q(x,y,t)=k \bar{p}(x,y,t)$ with Type 1 approach to obtain the constraints on the parameters of one-soliton solutions of the local equations (\ref{localcomplexAKNS(-3)}) and (\ref{localcomplexAKNS(-4)}). We have
\begin{equation}\label{relationonelocal}\displaystyle
\frac{e^{k_2x+\rho_2y+\omega_2t+\delta_2}}{1+Ae^{(k_1+k_2)x+(\rho_1+\rho_2)y+(\omega_1+\omega_2)t+\delta_1+\delta_2}}
=\frac{k e^{\bar{k}_1x+\bar{\rho}_1y+\bar{\omega}_1t+\bar{\delta}_1}}{1+\bar{A}e^{(\bar{k}_1+\bar{k}_2)x+(\bar{\rho}_1+\bar{\rho}_2)y+(\bar{\omega}_1+\bar{\omega}_2)t
+\bar{\delta}_1+\bar{\delta}_2}}.
\end{equation}

\noindent Here the constraints are obtained as
\begin{equation}
1)\,k_2=\bar{k}_1,\quad  2)\, \omega_2=\bar{\omega}_1,\quad  3)\, e^{\delta_2}=k e^{\bar{\delta}_1},
\end{equation}
for both $(2+1)$-dimensional local reduced complex AKNS(-3) and AKNS(-4) equations. Under these constraints, we get the relation $\rho_2=\bar{\rho}_1$
directly. Therefore one-soliton solutions of (\ref{localcomplexAKNS(-3)}) and (\ref{localcomplexAKNS(-4)}) are
given by
\begin{equation}\label{localsoln}
p(x,y,t)=\frac{e^{k_1x+\rho_1y+\omega_1t+\delta_1}}{1-\frac{k}{(k_1+\bar{k}_1)^2}
e^{(k_1+\bar{k}_1)x+(\rho_1+\bar{\rho}_1)y+(\omega_1+\bar{\omega}_1)t+\delta_1+\bar{\delta}_1}}.
\end{equation}
Here for $N=3$, $a$ is a real and $b$ is a pure imaginary number, and $\rho_1=\frac{1}{b}(\frac{1}{2}k_1\omega_1-\frac{a}{8}k_1^4)$. For $N=4$,
$a$ and $b$ are pure imaginary numbers, and $\rho_1=\frac{1}{b}(\frac{1}{2}k_1\omega_1-\frac{a}{16}k_1^5)$.

\noindent Let $k=-(k_1+\bar{k}_1)^2e^{\alpha}<0$, $\alpha$ is a real constant. Then the solution (\ref{localsoln}) can be written as
\begin{equation}
p(x,y,t)=\frac{e^{\psi}}{2 \cosh (\theta)},
\end{equation}
where
\begin{eqnarray}
&&\theta= \frac{1}{2}[(k_{1}+\bar{k}_{1}) x+(\rho_{1}+\bar{\rho}_{1}) y+(\omega_{1}+\bar{\omega}_{1}) t+\delta_{1}+\bar{\delta}_{1}+\alpha],\\
&&\psi = \frac{1}{2}[(k_{1}-\bar{k}_{1}) x+(\rho_{1}-\bar{\rho}_{1}) y+(\omega_{1}-\bar{\omega}_{1}) t+\delta_{1}-\bar{\delta}_{1}-\alpha].
\end{eqnarray}
The real-valued solution $|p(x,y,t)|^2$ is
\begin{equation}
|p(x,y,t)|^2=\frac{e^{-\alpha}}{4\cosh^2(\theta)}.
\end{equation}
Hence for $k<0$, one-soliton solutions of the local reduced complex equations (\ref{localcomplexAKNS(-3)}) and (\ref{localcomplexAKNS(-4)}) are nonsingular and bounded.

\subsection{One-soliton solutions of nonlocal reduced equations}

\noindent We have two types of nonlocal reductions for $(2+1)$-dimensional AKNS(-3) and AKNS(-4) systems. We consider each case separately. \\

\noindent \textbf{a)}\, $q(x,y,t)= k p(\epsilon_1x,\epsilon_2y,\epsilon_3t)$, $k$ is a real constant. From that relation, when we use the Type 1 approach we obtain the constraints on the parameters of the one-soliton solutions as
\begin{equation}
1)\, k_2=\varepsilon_1 k_1,\quad 2)\, \omega_2=\varepsilon_3 \omega_1,\quad 3)\, e^{\delta_2}=k e^{\delta_1}.
\end{equation}
Under these conditions, the relation $\rho_2=\epsilon_2 \rho_1$ holds directly.\\

\noindent For $N=3$, we have the case when $(\varepsilon_1,\varepsilon_2,\varepsilon_3)=(1,-1,1)$. Therefore we get one-soliton solution of the $(2+1)$-dimensional reduced nonlocal AKNS(-3) equation (\ref{realnonlocalredn=3}) as
\begin{equation}\label{realnonsoln=3-1}
p(x,y,t)=\frac{e^{k_1x+\rho_1y+\omega_1t+\delta_1}}{1-\frac{k}{4k_1^2}
e^{2k_1x+2\omega_1t+2\delta_1}},
\end{equation}
where $\rho_1=\frac{1}{b}(\frac{1}{2}k_1\omega_1-\frac{a}{8}k_1^4)$.  Let us take the parameters $k_1, \omega_1, \delta_1, a$, and $b$ real. Hence
$\rho_1$ is also real. Now let $k=-4 k_{1}^2 e^{2\alpha}<0$ where $\alpha$ is a real constant then the solution (\ref{realnonsoln=3-1}) can be written as
\begin{equation}\label{realnonsoln=3-2}
p(x,y,t)=\frac{e^{\psi}}{1+e^{2 \theta+2 \alpha}},
\end{equation}
where
\begin{eqnarray}
&&\psi=k_{1} x+\rho_{1} y+ \omega_{1} t+ \delta_{1}, \\
&&\theta=k_{1} x+\rho_{1} t+\delta_{1}.
\end{eqnarray}
The solution (\ref{realnonsoln=3-2}) can be rewritten as
\begin{equation}\label{realnonsoln=3-3}
p(x,y,t)=\frac{e^{\rho_{1} y-\alpha}}{2 \cosh (\theta+\alpha)}.
\end{equation}
Hence for $k<0$, one-soliton solution of the $(2+1)$-dimensional reduced nonlocal AKNS(-3) equation (\ref{realnonlocalredn=3}) is nonsingular for all $(x,y,t)$ but unbounded.\\

\noindent For $N=4$, we have the case when $(\varepsilon_1,\varepsilon_2,\varepsilon_3)=(1,1,-1)$. One-soliton solution of the $(2+1)$-dimensional reduced nonlocal AKNS(-4) equation (\ref{realnonlocalredn=4}) is
\begin{equation}\label{realnonsoln=4-1}
p(x,y,t)=\frac{e^{k_1x+\rho_1y+\omega_1t+\delta_1}}{1-\frac{k}{4k_1^2},
e^{2k_1x+2\rho_1y+2\delta_1}},
\end{equation}
where $\rho_1=\frac{1}{b}(\frac{1}{2}k_1\omega_1-\frac{a}{16}k_1^5)$. Let us assume that all the parameters that $k_1, \omega_1, \delta_1, a$, and $b$ are real.
In that case $\rho_1$ is also real. Let now $k=-4 k_{1}^2 e^{2\alpha}<0$ where $\alpha$ is a real constant then the solution (\ref{realnonsoln=4-1}) becomes
\begin{equation}\label{realnonsoln=4-1}
p(x,y,t)=\frac{e^{\psi}}{1+e^{2 \theta+2 \alpha}},
\end{equation}
where
\begin{eqnarray}
&&\psi=k_{1} x+\rho_{1} y+ \omega_{1} t+ \delta_{1}, \\
&&\theta=k_{1} x+\rho_{1} y+\delta_{1}.
\end{eqnarray}
The solution (\ref{realnonsoln=4-1}) can be rewritten as
\begin{equation}\label{realnonsoln=4-2}
p(x, y, t)=\frac{e^{\omega_{1} t-\alpha}}{2 \cosh (\theta+\alpha)}.
\end{equation}
Therefore for $k<0$, one-soliton solution is bounded for all $t \ge 0$ and $\omega_1\leq 0$, and nonsingular for all $(x,y,t)$.\\

\noindent For both nonlocal reduced AKNS(-3) and AKNS(-4) equations we also have the case when $(\epsilon_1,\epsilon_2,\epsilon_3)=(-1,-1,-1)$. If we use the Type 1 approach for this case we obtain trivial solution $p(x,y,t)=0$. Therefore here we use Type 2 approach \cite{GurPek2}, \cite{GurPek3}, which is based on the cross multiplication. From the reduction formula we have
\begin{equation}\displaystyle
\frac{e^{\theta_2}}{1+Ae^{\theta_1+\theta_2}}=k\frac{e^{\theta_1^{-}}}{1+Ae^{\theta_1^{-}+\theta_2^{-}}},
\end{equation}
where
\begin{equation*}
\theta_j=k_jx+\rho_j y+\omega_jt+\delta_j,\quad \theta_j^{-}=-k_jx-\rho_j y-\omega_jt+\delta_j,\quad j=1, 2.
\end{equation*}
By the cross multiplication we get
\begin{equation}
e^{\theta_2}+Ae^{2\delta_2}e^{\theta_1^{-}}=ke^{\theta_1^{-}}+Ake^{2\delta_1}e^{\theta_2},
\end{equation}
yielding the conditions
\begin{equation}
 Ak e^{2\delta_1}=1,\quad \quad Ae^{2\delta_2}=k.
\end{equation}
Hence we get
\begin{equation}
 e^{\delta_1}=\sigma_1 i\frac{(k_1+k_2)}{\sqrt{k}}, \quad e^{\delta_2}=\sigma_2 i\sqrt{k}(k_1+k_2),\quad \sigma_j=\pm 1,\,\, j=1, 2.
\end{equation}
\noindent Therefore one-soliton solutions of the nonlocal equations (\ref{realnonlocalredn=3}) and (\ref{realnonlocalredn=4}) for $p^{\varepsilon}=p(-x,-y,-t)$ are given by
\begin{equation}\label{type2sol-1}\displaystyle
p(x,y,t)=\frac{i\sigma_1 e^{k_1x+\rho_1y+\omega_1t}(k_1+k_2)}{\sqrt{k}(1+\sigma_1\sigma_2e^{(k_1+k_2)x+(\rho_1+\rho_2)y+(\omega_1+\omega_2)t})}, \quad \sigma_j=\pm 1, j=1, 2,
\end{equation}
where $\rho_1$ and $\rho_2$ satisfy the dispersion relation (\ref{dispersionAKNS(-3)}) for $N=3$ and the relation (\ref{dispersionAKNS(-4)}) for $N=4$. We can rewrite the solution
(\ref{type2sol-1}) as
\begin{equation}\label{type2sol-2}\displaystyle
p(x,y,t)=\frac{e^{\psi+\delta_1}}{2\cosh(\theta)},
\end{equation}
where
\begin{align}
&\psi=\frac{1}{2}[(k_1-k_2)x+(\rho_1-\rho_2)y+(\omega_1-\omega_2)t],\\
&\theta=\frac{1}{2}[(k_1+k_2)x+(\rho_1+\rho_2)y+(\omega_1+\omega_2)t].
\end{align}
Let all the parameters be real. Then the solution (\ref{type2sol-2}) for nonlocal reduced AKNS(-3) equation (\ref{realnonlocalredn=3})
is nonsingular and bounded if $k_1=k_2$, $\rho_1=\rho_2$ yielding $\omega_2=\frac{a}{2}k_1^3-\omega_1$ and $\omega_1\leq \omega_2$ for
$t\geq 0$. The solution (\ref{type2sol-2}) for nonlocal reduced AKNS(-4) equation (\ref{realnonlocalredn=4})
is nonsingular and bounded if $k_1=k_2$, $\rho_1=\rho_2$ yielding $\omega_2=-\omega_1$ and $\omega_1\leq 0$ for $t\geq 0$.\\

\noindent \textbf{b)}\, $q(x,y,t)= k \bar{p}(\epsilon_1x,\epsilon_2y,\epsilon_3t)$, $k$ is a real constant. When we use Type 1, we obtain the constraints for the parameters of one-soliton solutions of the nonlocal reduced complex AKNS(-3) and AKNS(-4) equations as
\begin{equation}
1)\, k_2 = \varepsilon_1\bar{k}_1,\quad 2)\, \omega_2=\varepsilon_3\bar{\omega}_1, \quad  3)\, e^{\delta_2}=ke^{\bar{\delta}_1},
\end{equation}
which are same as in the cases of AKNS(0), AKNS(-1), and AKNS(-2) systems \cite{GurPek4}. The above relations yield $\rho_2=\varepsilon_2\bar{\rho}_1$
directly. Hence one-soliton solutions of the nonlocal reduced complex AKNS(-3) and AKNS(-4) equations given by (\ref{complexnonlocalredn=3}) and (\ref{complexnonlocalredn=4}) respectively, are
\begin{equation}\label{nonlocalsoln2}
p(x,y,t)=\frac{e^{k_1x+\rho_1y+\omega_1t+\delta_1}}{1-\frac{k}{(k_1+\varepsilon_1\bar{k}_1)^2}
e^{(k_1+\varepsilon_1\bar{k}_1)x+(\rho_1+\varepsilon_2\bar{\rho}_1)y+(\omega_1+\varepsilon_3\bar{\omega}_1)t+\delta_1+\bar{\delta}_1}},
\end{equation}
with the corresponding dispersion relations.\\

\noindent One can find nonsingular and bounded solutions for different choices of $(\varepsilon_1, \varepsilon_2, \varepsilon_3)$. There are 14 nonlocal complex reduced equations obtained by the reduction $q(x,y,t)= k \bar{p}(\epsilon_1x,\epsilon_2y,\epsilon_3t)$. Let us consider $x$-reversal case that is when
$(\varepsilon_1, \varepsilon_2, \varepsilon_3)=(-1,1,1)$. Let also $k_1=\alpha_1+i\beta_1$, $\tau_1=\alpha_2+i\beta_2$, $\omega_1=\alpha_3+i\beta_3$,
and $e^{\delta_1}=\alpha_4+i\beta_4$, where $\alpha_j, \beta_j \in \mathbb{R}$ for $j=1,2,3,4$. Then the solution (\ref{nonlocalsoln2}) for both nonlocal reduced complex AKNS(-3) and AKNS(-4) equations given by (\ref{complexnonlocalredn=3}) and (\ref{complexnonlocalredn=4}) for $(\varepsilon_1, \varepsilon_2, \varepsilon_3)=(-1,1,1)$ becomes
\begin{equation}
p(x,y,t)=\frac{e^{(\alpha_1+i\beta_1)x+(\alpha_2+i\beta_2)y+(\alpha_3+i\beta_3)t}(\alpha_4+i\beta_4)}
{1+\frac{k(\alpha_4^2+\beta_4^2)}{4\beta_1^2}e^{2i\beta_1x+2\alpha_2y+2\alpha_3t}}
\end{equation}
so
\begin{equation}
|p(x,y,t)|^2=\frac{e^{2\alpha_1x+2\alpha_2y+2\alpha_3t} }{(\frac{k(\alpha_4^2+\beta_4^2)}{4\beta_1^2}e^{2\alpha_2y+2\alpha_3t} +\cos(2\beta_1x))^2+\sin^2(2\beta_1x)}.
\end{equation}
For $x=\frac{n\pi}{2\beta_1}$ and $\frac{k(\alpha_4^2+\beta_4^2)}{4\beta_1^2}e^{2\alpha_2y+2\alpha_3t}+(-1)^n=0$, $n$ integer, the above solution is singular.

\section{Conclusion}
In this work we constructed the Hirota bilinear forms of the $(2+1)$-dimensional systems AKNS(-3) and AKNS(-4) and found one-soliton solutions of these systems. We then found all possible local and nonlocal reductions of these systems of equations. By using these solutions with the reduction formulas we presented one-soliton solutions of all reduced nonlocal equations.

\section{Acknowledgment}
  This work is partially supported by the Scientific
and Technological Research Council of Turkey (T\"{U}B\.{I}TAK).\\

\end{document}